\documentclass{emulateapj}
\usepackage{apjfonts}
\usepackage{amsmath}

\newcommand{\myemail}{kmurase@yukawa.kyoto-u.ac.jp}

\def\mr#1{\mathrm{#1}}

\newcounter{ichi}
\newcounter{ni}
\newcounter{san}
\slugcomment{Submitted}

\shorttitle{Cosmic Rays and Neutrinos from Clusters of Galaxies}
\shortauthors{Murase, Inoue, \& Nagataki}

\begin{document}


\title{
Cosmic Rays Above the Second Knee from Clusters of Galaxies\\
and Associated High-Energy Neutrino Emission}


\author{Kohta Murase$^{1}$}
\author{Susumu Inoue$^{2,3}$}
\author{Shigehiro Nagataki$^{1}$}%


\altaffiltext{1}{YITP, Kyoto University, Kyoto, Oiwake-cho,
Kitashirakawa, Sakyo-ku, Kyoto 606-8502, Japan; \myemail}
\altaffiltext{2}{Division of Theoretical Astronomy,
National Astronomical Observatory of Japan,
Mitaka, Tokyo 181-8588, Japan}
\altaffiltext{3}{Department of Physics, Kyoto University, Kyoto, Oiwake-cho,
Kitashirakawa, Sakyo-ku, Kyoto 606-8502, Japan}


\begin{abstract}
Accretion and merger shocks in clusters of galaxies are potential accelerators of high-energy protons,
which can give rise to high-energy neutrinos through $pp$ interactions with the intracluster gas.
We discuss the possibility that protons from cluster shocks make a significant contribution
to the observed cosmic rays in the energy range between the second knee at $\sim {10}^{17.5}$ eV
and the ankle at $\sim {10}^{18.5}$ eV.
The accompanying cumulative neutrino background above $\sim$PeV 
may be detectable by upcoming neutrino telescopes such as IceCube or KM3NeT,
providing a test of this scenario as well as a probe of cosmic-ray confinement properties in clusters.
\end{abstract}





\section{\label{sec:level1}Introduction}
Clusters of galaxies (CGs) represent the largest gravitationally bound objects in the universe \cite[e.g.,][]{V05}.
According to standard, hierarchical scenarios of cosmological structure formation,
they are the latest systems to virialize and continue to grow
through merging and accretion of dark matter and baryonic gas,
thereby generating powerful shock waves on Mpc scales.
In particular, accretion shocks with high Mach numbers are expected on the outskirts of massive CGs,
potentially leading to efficient acceleration of high-energy particles \citep[e.g.,][]{Mea00,Rea03}.
Here ``accretion'' signifies not only infall of diffuse intergalactic gas,
but also minor merger events that induce sufficiently strong shocks near the virial radii.
Moderate Mach number shocks arising further inside the CG could also be
important in certain situations \citep[e.g.,][]{Rea03}.
Predictions for the associated nonthermal radiation, notably high-energy gamma-rays,
have been discussed by a number of authors \citep[e.g.,][]{VAB96,BBP97,CB98,LW00,IAS05,AN08}.

Cosmic rays (CRs) are observed over 11 decades of energy
from $\sim 10^9$ eV to $\gtrsim 10^{20}$ eV,
and their origin is under intense debate.
The all-particle spectrum is characterized by broken power-laws with a number of breaks:
the knee at $\sim 10^{15.5}$ eV where the spectral index changes from $\sim -2.7$ to $\sim -3.0$,
the second knee at $\sim 10^{17.5}$ eV where it changes from $\sim -3.0$ to $-3.2$,
and the ankle at $\sim 10^{18.5}$ eV where it changes from $\sim -3.2$ to $\sim -2.7$ \cite{NW00}.
Galactic supernova remnants (SNRs) are widely believed
to be responsible for CRs at least up to the knee,
and probably up to somewhat higher energies 
\citep[and references therein]{H05}.
In contrast, ultra-high-energy CRs (UHECRs) with energies above the ankle
are generally thought to be extragalactic \citep[e.g.,][]{NW00,GS06,BB07}.
The Pierre Auger Observatory has revealed an anisotropy
in the arrival directions of UHECRs that supports this view 
(Abraham et al. 2007; 2008; see however Abbasi et al. 2008)
but the actual identity of the sources remain unknown \citep[e.g.,][]{K08,I08}.
The origin of CRs with energies between the second knee and the ankle is even more controversial,
and both extragalactic and Galactic viewpoints have been advocated.
The sources may be the same as those for UHECRs \cite{BGG06,Aea07},
or something entirely different such as 
specific types of supernovae, 
hypernovae, Galactic winds, etc \citep[][and references there in]{GS06,BB07,WRMD07,Bea08}.


In this letter, we discuss high-energy neutrino emission from CGs
in light of the possibility that accretion and/or merger shocks in CGs
are important sources of CRs with energies between the second knee and the ankle.
The observability by near future telescopes such as IceCube \cite{Aea04} or KM3NeT \cite{K06} is addressed.
Note that this view is different from scenarios in which 
cluster shocks are the main sources of UHECRs above the ankle \cite{NMA95,KRJ96,KRB97,ISMA07}.
More detailed discussions including the accompanying hadronic gamma-ray emission
will be presented in a subsequent paper.
We adopt the $\Lambda$CDM cosmological parameters
$\Omega_b=0.04$, $\Omega_m=0.3$, $\Omega_{\Lambda}=0.7$ and $h=0.7$.

\section{Cosmic-Ray Production}
We first estimate the maximum energy of the accelerated CRs.
The virial radius of a CG with mass $M=10^{15} M_{15} {\rm M_\odot}$ is
$r_{\rm{vir}} \simeq 2.4 \, {\rm{Mpc}}\, M_{15}^{1/3} F(z,\Omega_m) {(h/0.7)}^{-1} {(1+z)}^{-1}$,
where $F(z,\Omega_m)$ is a factor of order unity that depends weakly on redshift $z$ and $\Omega_m$ \cite{V05}.
We write the shock radius as $r_{\rm{sh}} \equiv \lambda_{\rm{sh}} r_{\rm{vir}}$,
with $\lambda_{\rm{sh}} \sim 1-10$ expected for accretion shocks \citep[e.g.,][]{Rea03}.
The typical shock velocity $V_{\rm{sh}}$ should be comparable to the velocity of the infalling gas
$V_{\rm{ff}} \simeq 2000 \, {\rm{km} \, \rm{s}^{-1}} \, M_{15}^{1/2} 
{(r_{\rm{sh}}/1 \, \rm{Mpc})}^{-1/2}$ \cite{IAS05}.

The typical shock acceleration time for CRs with energy $\varepsilon$ and charge $Z$ is
$t_{\rm{acc}} \approx 20 \kappa_{\rm{sh}}/V_{\rm{sh}}^2=(20/3) (c \varepsilon /Z e B V_{\rm{sh}}^2)~\xi$
\citep[see, e.g.,][for reviews]{BE87}.
Here $B$ and $\kappa_{\rm{sh}}$ are respectively the magnetic field and diffusion coefficient at the shock,
and $\xi \equiv {(B/\delta B)}^{2}$ where $\xi \rightarrow 1$ in the Bohm limit.
Although the magnetic fields at cluster shocks are uncertain,
we take $B \sim 1 \, \mu$G, as supported by recent X-ray observations
of diffuse radio relics near $r_{\rm{vir}}$ for several CGs \cite{FN06,Hea08,Nea08}.
We also postulate $\xi \sim 1$, as observed to be the case for some SNRs \cite{Uea07},
and which can be expected if the fields are generated locally at the shocks
by mechanisms such as the CR streaming instability \cite[see, e.g.,][]{BL01,B04,VEB06,BDD08}.

The maximum energy of the accelerated CRs $\varepsilon_{\max}$ can be estimated
by equating $t_{\rm{acc}}$ with various limiting time scales,
such as the diffusive escape time from the shock
$t_{\rm{esc}} \approx r_{\rm{sh}}^2/6\kappa_{\rm{sh}}$,
and the energy loss time due to photohadronic and/or photodisintegration interactions 
with the CMB and the infrared (IR) background \cite{KRB97,ISMA07}.
When the shock is due to a transient merger-like event,
the lifetime of the shock may also be relevant,
which can be estimated by the dynamical time $t_{\rm{dyn}} = r_{\rm{sh}}/V_{\rm{sh}}$.
In the latter case, $\varepsilon_{\max}$ would generally be determined by $t_{\rm{dyn}}$ so that
${\varepsilon}_{\max} \simeq 1.6 \times 10^{18} \, {\rm{eV}} \, Z {\xi}^{-1} M_{15}^{2/3} B_{-6}
\lambda_{\rm{sh}}^{1/2}$, where $B= B_{-6} \mu$G. Hence, we expect
that cluster shocks can accelerate at least protons up to around the ankle \cite{NMA95}.

Next we consider the energetics.
Assuming a total mass accretion rate $\dot{M} \approx 0.1 V_{\rm{ff}}^3/G$
and gas fraction $f_g =\Omega_b/\Omega_m \sim 0.13$,
the dissipation rate of infalling gas kinetic energy through the accretion shock
of a CG with mass $M$ is estimated to be
$L_{\rm{ac}} \approx f_g G M \dot{M}/r_{\rm{sh}} \simeq 7 \times
10^{45} {\rm{erg}} \, {\rm{s}^{-1}} \, (f_g/0.13) M_{15}^{5/3}$ \cite{KWL04}.
Taking the local density of massive CGs $n_{\rm{CG}} (M \gtrsim M_{15}) 
\sim 2 \times 10^{-6} \, \rm{Mpc}^{-3}$ \cite{Jea01}
and a CR injection efficiency $\epsilon_{\rm{acc}} \sim 0.2$,
the CR power from CGs per logarithmic energy interval at $\varepsilon =10^{18} \, {\rm{eV}}$ is
$\varepsilon^2 (d \dot{n}/d\varepsilon)
\simeq {10}^{45}\, {\rm{erg} \, \rm{Mpc}^{-3} \rm{yr}^{-1}} \, {(R/50)}^{-1}$.
Here $R(\varepsilon) \equiv
\left(\int_{\varepsilon_{\min}}^{\varepsilon_{\max}} d \varepsilon^{\prime} \, 
\varepsilon^{\prime} (d N/d\varepsilon^{\prime}) \right)
/ (\varepsilon^2 d N/d\varepsilon)$
depends on the injection CR spectrum;
in the case of a single power-law with index $p$ and minimum energy $\varepsilon_{\min} = 1$ GeV,
$R \sim 25$ for $p=2.0$, and $R \sim 300$ for $p=2.2$ \cite{MISN08}.

In comparison, the observed CR spectrum for $10^{17}$ eV $<\varepsilon<$ $10^{18.5}$ eV
is $\Phi \simeq 9.23 \times {10}^{-28} \, {\rm{eV}^{-1} \rm{cm}^{-2} \rm{s}^{-1} \rm{sr}^{-1}} \, 
{(\varepsilon/6.3 \times {10}^{18} \, \rm{eV})}^{-3.2}$ \cite{NW00}.
The implied CR source power and spectrum depend on evolution effects
and the uncertain CR composition at these energies .
Here we assume that they are proton-dominant as in some extragalactic scenarios \cite{BGG06},
and cases with a more general composition will be discussed in the future work.
The required CR power at $\varepsilon =10^{18} \, {\rm{eV}}$ is then
$\varepsilon^2 (d \dot{n}/d\varepsilon) \sim 
{10}^{45} \, \rm{erg} \, \rm{Mpc}^{-3} \rm{yr}^{-1}$,
crudely accounting for energy losses during diffusive intergalactic propagation \cite{BGG06}.
As long as $R \lesssim 100$, the two powers would be comparable and CGs energetically viable.
However, since intergalactic propagation should steepen
the spectral index from the injection value $p$ by $\sim 0.5-0.8$ \cite{BGG06},
the observed index of -3.2 requires $p \sim 2.4-2.7$ 
and hence $R \gg 100$ for single power-law spectra extending down 
to $\varepsilon_{\min}=1$ GeV.
Avoiding excessive energy demands
motivates a broken power-law form with break energy $\varepsilon_b$:
$dN/d \varepsilon \propto {\varepsilon}^{-p_1}$
for $\varepsilon < {\varepsilon}_b$ and
$dN/d \varepsilon \propto
{\varepsilon}^{-p_2} \exp(-(\varepsilon/\varepsilon_{\max}))$ for 
$\varepsilon \geq {\varepsilon}_b$ \cite[see also][]{Aea07}.

We consider two possibilities as to how such spectra may actually occur.
One is through the superposition of hard spectra ($p_1 \sim 2.0$) 
with a distribution of $\varepsilon_{\max}$ \cite{KS06},
which can be related to accretion shocks with a distribution of $M$.
It was seen that $\varepsilon_{\max} \propto M^{2/3}$
if the relevant condition is $t_{\rm acc} \approx t_{\rm dyn}$
and if the $M$-dependence of $B$ is not strong.
A realistic CG mass function $n_{\rm{CG}}(>M) \propto M^{-1} \exp(-(M/ 1.8 \times {10}^{14} M_{\odot}))$
can be approximated over a limited range of $M$ as a power-law $n_{\rm{CG}}(>M) \propto M^{-\alpha}$ \cite{Jea01}
so that $dN/d\varepsilon \propto {\varepsilon}^{-\frac{3 \alpha}{2}-\frac{1}{3}}$ for $\varepsilon > \varepsilon_b$.
This could allow $p_1 \sim 2.0$ and $p_2 \sim 2.0-3.3$.

Another possibility is a two-step acceleration process,
a first source providing a seed CR population with hard spectra ($p_1
\sim 2.0$) up to $\varepsilon_b$,
which is then picked up by a second source and accelerated further
with softer spectra to $\varepsilon_{\max}$.
Since CRs with sufficiently low energies are likely to be confined in the intracluster medium (ICM)
for very long times \cite{VAB96,BBP97},
the seed population can come from a number of sources, all accumulated over the history of the CG:
the low energy portion of accretion shock CRs,
supernova-driven galactic winds (GWs),
and the jets of radio-loud active galactic nuclei (AGNs).
Their relative importance can be estimated through their contributions to the heating of the ICM,
which should be roughly proportional to their CR output as long as the relevant shocks are sufficiently strong.
In the absence of GWs or AGNs,
high Mach number accretion shocks are expected to contribute $\sim$10\% of the heating of the ICM,
while the remainder is mediated by low Mach number merger shocks \citep[e.g.,][]{Rea03}.
GWs are unlikely to play a significant role in ICM heating
due to severe radiative losses during their formation \cite[e.g.][]{KY00}.
(Note that CRs of supernova origin escaping from within the host galaxy will also suffer heavy adiabatic losses).
In contrast, AGN jets can contribute 1-2 keV per baryon of heat input directly to the ICM \cite{IS01}
\cite[see also][]{Eea97,Eea98}.
For massive clusters with temperatures $\sim$10 keV,
this implies that CRs from AGNs can be energetically comparable to those from accretion shocks,
and may be even higher for less massive clusters.
Subsequent acceleration of these seed CRs to $\varepsilon_{\max}$ may be achieved
through merger and/or accretion shocks with moderate Mach numbers 
$\mathcal{M} \sim 2.5-5$, leading to $p_2 \sim 2.2-2.7$. 
The break energy $\varepsilon_{b}$ may correspond to
the confinement energy $\varepsilon_{\rm{diff}}$
above which CRs begin to escape diffusively out of the ICM.
Under Kolmogorov-like turbulence, ${\varepsilon}_{\rm{diff}}
\simeq 1.8 \times {10}^{17} \, {\rm{eV}} \, Z 
{(r/1.5 \, \rm{Mpc})}^{6} \kappa_{\rm{CG},30}^{-3}
{(\Delta t/1 \, \rm{Gyr})}^{-3}$,
where $\kappa_{\rm{CG}} = 10^{30} \kappa_{\rm{CG},30} {\rm \ cm^2
{s}^{-1}}$ is the diffusion coefficient in the ICM at $\varepsilon=1$ GeV,
and $\Delta t$ is the time elapsed after injection \cite{VAB96,BBP97}.
Within the uncertainties, we see that CGs could be a viable source of CRs
with energies between the second knee and the ankle.

\section{Neutrino Production}
We now evaluate the spectra of associated gamma-rays and neutrinos,
which are inevitably generated through $pp$ interactions with the ambient ICM gas \cite{VAB96,BBP97,CB98}.
In view of the above, we assume that CRs with a broken power-law spectrum is realized
with $p_1=2.0$ and $p_2=2.4$. We choose $\varepsilon_b={10}^{16.5}$ eV or ${10}^{17.5}$ eV,
giving respectively $R \simeq 78$ or $35$.
The spatial distribution of the thermal ICM gas is
generally well-constrained from X-ray observations \cite{PE04}.
However, that for the CRs is uncertain, and we consider the following four models.
\textit{Model A}: CRs are uniformly distributed within $r_{\rm sh}$,
with $\lambda_{\rm{sh}}$ chosen such that $t_{\rm{dyn}}=1$ Gyr.
\textit{Model B}: CRs are uniformly distributed within $r_{\rm vir}$,
giving a conservative estimate compared to other models.
\textit{Isobaric}: CRs at each radii have energy density
proportional to that of the thermal gas with ratio $X_{\rm{CR}}$ \cite{PE04,AN08}.
\textit{Central AGN}: CRs are distributed as
$dN/d \varepsilon \propto r^{-1} {\varepsilon}^{-p-\frac{1}{3}}$ 
for $\varepsilon \geq {(r_c^2/6 \kappa_{\rm{CG}} \Delta t)}^{3}$
and $dN/d \varepsilon \propto {\varepsilon}^{-p}$ for $\varepsilon < 
{(r_c^2/6 \kappa_{\rm{CG}} \Delta t)}^{3}$,
corresponding to CRs diffusing out from a central source such as an AGN
as discussed in \cite{BBP97,CB98};
here accretion or merger shocks may not be involved. 
But, in the two-step acceleration scenario, this model could be more realistic
below the break energy.
We perform numerical calculations of the neutrino spectra 
using formulae based on the SIBYLL code at high energies \cite{KAB06}.

\begin{figure}[t]
\includegraphics[width=\linewidth]{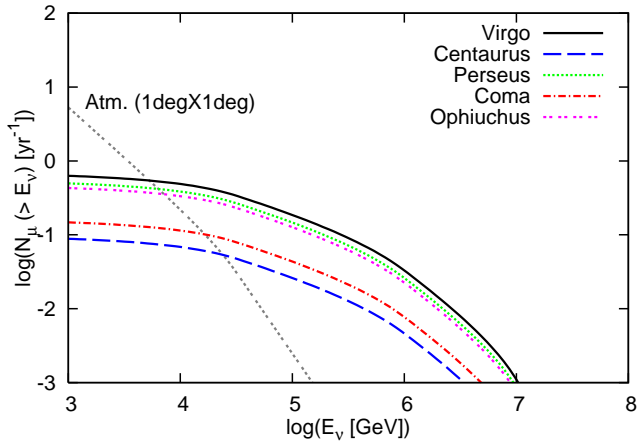}
\caption{\small{\label{Fig1} Expected event rates for
muon neutrinos ($\nu_{\mu} + \bar{\nu}_{\mu}$) in IceCube-like
detectors from five nearby CGs: Virgo, Centaurus, Perseus, Coma and Ophiuchus.
Broken power-law CR spectra with $p_1=2.0$, $p_2=2.4$ and $\varepsilon_b={10}^{17.5}$ eV
is assumed, and the \textit{Isobaric} model with $X_{\rm{CR}}=0.029$ is used.
Note that IceCube and KM3NeT mainly cover the northern and southern celestial hemispheres,
respectively. Neutrino oscillation is taken into account.}}
\end{figure}
\begin{figure}[t]
\includegraphics[width=\linewidth]{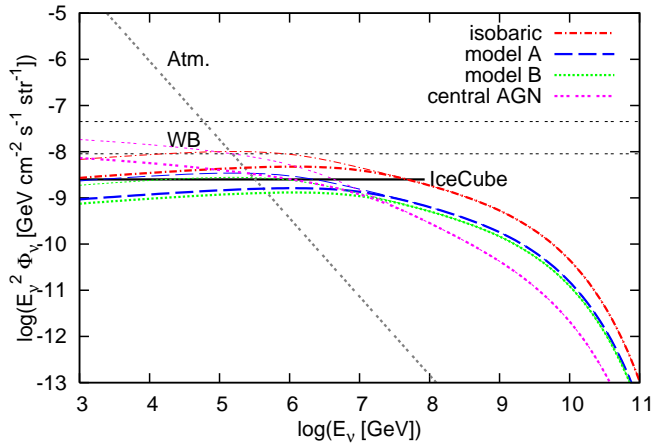}
\caption{\small{\label{Fig2} Cumulative neutrino ($\nu_{e} + 
\bar{\nu}_{e} + \nu_{\mu} + \bar{\nu}_{\mu} + \nu_{\tau} + 
\bar{\nu}_{\tau}$) background from CGs for broken
power-law CR spectra with $p_1=2.0$ and $p_2=2.4$. 
The break energies are $\varepsilon_b={10}^{17.5}$ eV (thick lines) and 
$\varepsilon_b={10}^{16.5}$ eV (thin lines), respectively.  
The CR power is normalized to 
$\varepsilon^2 (d \dot{n}/d \varepsilon) = 2 \times {10}^{45} \, \rm{erg} \, 
\rm{Mpc}^{-3} \rm{yr}^{-1}$ at $\varepsilon={10}^{18}$ eV,
as required to account for CRs above the second knee. For \textit{isobaric},
the corresponding $X_{\rm{CR}}$ is 0.029 and 0.067. For \textit{central AGN},
Kolmogorov-like turbulence is assumed with
$\kappa_{\rm{CG}}={10}^{30}\, \rm{cm}^{2} {s}^{-1}$. We take 
$t_{\rm{dyn}}= \Delta t = 1$ Gyr and $z_{\max}=2$.  
WB represents the Waxman-Bahcall bounds \cite{WB98}.}}
\end{figure}

The neutrino and gamma-ray fluxes can be estimated
analytically via the effective optical depth for the $pp$ reaction
as $f_{pp} \approx 0.8 \sigma_{pp} n_{\rm{N}} c t_{\rm{int}}$,
where $n_{\rm{N}}$ is the target nucleon
density in the ICM, $\sigma_{pp}$ is the $pp$ cross section and
$t_{\rm{int}} \sim t_{\rm dyn}$ or ${\rm{max}}[r/c,t_{\rm{diff}}]$ is the
$pp$ interaction time.
Because $n_{\rm{N}} \sim {10}^{-4.5} \, \rm{cm}^{-3}$ at $r \sim 1.5$
Mpc \cite{CB98,PE04} and $\sigma_{pp} \sim {10}^{-25} \, \rm{cm}^{2}$ in the $100$ PeV range \cite{KAB06},
we obtain
\begin{eqnarray}
f_{pp} &\sim& 2.4 \times {10}^{-3} n_{\rm{N},-4.5} 
(t_{\rm{int}}/1 \, \rm{Gyr}). \label{pp}
\end{eqnarray} 
Roughly speaking, high-energy neutrinos from charged-pion decay have
typical energy $\varepsilon_{\nu} \sim 0.03 \varepsilon$
(true only in the average sense, because charged particles have wide
energy distributions
and high multiplicities as expected from the KNO scaling law) \cite{KAB06}.
Hence, neutrinos $\gtrsim$ PeV are directly related to CRs above the
second knee.

First we obtain numerically the neutrino spectra and expected event rates from five nearby CGs,
utilizing the $\beta$ model or double-$\beta$ model description
in Tables. 1 and 2 in Pfrommer \& En{\ss}lin 2004 for the thermal gas profile of each CG (Fig. 1).
Our gamma-ray fluxes for single power-law spectra agree with the results of Pfrommer \& En{\ss}lin 2004.
As is apparent in Fig. 1, the detection of neutrino signals from 
individual CGs could be challenging even for nearby objects.
It may be achievable, however, through a detailed stacking analysis.

More promising would be the cumulative background signal.
A rough estimate of the neutrino background is \citep[e.g.,][]{M07,WB98}
\begin{eqnarray}
\varepsilon_{\nu}^2 \Phi_{\nu} &\sim& 
\frac{c}{4\pi H_{0}} \frac{1}{3} \mr{min}[1,f_{pp}] 
\varepsilon^2 \frac{d N}{d\varepsilon dt} n_{\mr{CG}}(0) f_{z} \nonumber\\
&\sim&  1.5 \times 10^{-9} \, {\rm{GeV cm^{-2} s^{-1} str^{-1}}} \,
f_{z} \nonumber\\
&\times& \left(\frac{f_{pp}(\varepsilon={10}^{18} \, \rm{eV})}
{2.4 \times {10}^{-3}}\right)
{\left(\frac{\varepsilon_{\nu}}{10 \, \rm{PeV}}  \right)}^{-p+2.1}, \label{bkg}
\end{eqnarray}
where CGs are assumed to be the main sources of CRs from the second knee to the ankle.
Here, $n_{\rm{CG}}(0)$ is the local density of massive CGs
and $f_z$ is a correction factor for the source evolution \cite{M07,WB98}.
For detailed numerical calculations of the background,
we treat more distant CGs following \cite{CB98}, adopting the mass function of \cite{Jea01}.  
The results for the broken power-law case are shown in Fig. 2.
With $\varepsilon_b = {10}^{17.5}$ eV, 
the expected event rates above 0.1 PeV in IceCube \cite{Aea04}
are $\sim2 \ {\rm yr^{-1}}$ for \textit{model A}, $\sim1 \ {\rm yr^{-1}}$ for \textit{model B},
$\sim5 \ {\rm yr^{-1}}$ for \textit{isobaric} and $\sim 3 \ {\rm yr^{-1}}$ for \textit{central AGN}.

Hence, upcoming telescopes may be able to find multi-PeV neutrino signals from CGs,
providing a crucial test of our scenario.
From Eq. (\ref{bkg}), we can also estimate the corresponding gamma-ray background
from $\pi^0$ decay, which is $\varepsilon_{\gamma}^2 \Phi_{\gamma} \sim 
({10}^{-9}-{10}^{-8}) \, \rm{GeV} {cm}^{-2} {s}^{-1} {str}^{-1}$ for
the broken power-law case. 
This is only $(0.1-1)$\% of the EGRET limit,
consistent with the nondetection so far for individual CGs.
Note that the expected gamma-ray background flux would increase
if $\varepsilon_b$ can be decreased, requiring larger CR power from CGs.

\section{Implications and Discussion}
To test the CG origin of second knee CRs,
high-energy neutrinos should offer one of the most crucial multi-messenger signals.
Unlike at the highest energies, CRs themselves in the $10^{18}$ eV range
offer no chance of source identification
as they should be severely deflected by Galactic and extragalactic
magnetic fields. Moreover, due to magnetic horizon effects,
extragalactic CRs $\lesssim 10^{17}$ eV may not reach us at all \cite{L05,KL07},
so even the broken power-law spectral form will not be directly
observable. Gamma-rays are unaffected by intervening magnetic fields,
but those at $\gtrsim$ PeV energies relevant for the second knee
are significantly attenuated by pair-creation processes
with the CMB and cosmic IR backgrounds \citep[e.g.,][]{K08}.
In constrast, neutrinos in the PeV-EeV energy range should be unscathed during propagation
\citep[][and references there in]{BS00}. 
Consequently, such neutrinos may also constitute a unique tool
for probing the uncertain CR confinement properties of CGs
through the dependence on $\varepsilon_{\rm diff}$.

AGNs can complicate the cluster shock neutrino signal,
either by emitting PeV-EeV neutrinos themselves,
or injecting CRs that produces neutrinos without the intervention of cluster shocks.
In principle, cross correlation of the detected events with known CGs and AGNs
should be an effective discriminant.
In the former case, AGNs inside as well as outside CGs should correlate,
whereas in the latter, CGs with powerful AGNs should correlate stronger than those without.
For cluster shock neutrinos, correlation with all sufficiently massive CGs is expected.


Gamma-ray observations at GeV-TeV energies would also be crucial.
In combination with $\gtrsim$ PeV neutrino observations,
they can probe the CR spectrum over a broad energy range
and test our broken power-law assumption.
By providing information on the spatial distribution of sub-PeV CRs,
they would also help to distinguish among our different models,
and to constrain the AGN contribution of CRs and neutrinos.
However, if $\varepsilon_b$ is sufficiently high, their detection may not be
trivial except for a few nearby CGs such as Virgo.
Other emission processes may also be at work \citep[e.g.,][]{LW00,IAS05},
complicating the extraction of the $\pi^0$ decay component.
More details on the gamma-ray emission will be given in a following paper. 

Note that high-energy neutrinos can also be produced by photomeson interactions
with the IR background \citep[e.g.,][]{TMNS07}, which may possibly be enhanced 
inside CGs \cite{MHSB06}. The effective optical depth for this process
is roughly $f_{p \gamma} \sim {10}^{-3}-{10}^{-1}$.
These neutrinos become important above $\sim 0.1$ EeV,
where detection by IceCube or KM3NeT is relatively more difficult.
We defer the study of such photomeson-induced neutrinos in CGs to the future.

Our neutrino predictions were based on simplified models for the CR distribution in CGs.
More realistic evaluations need to include the inhomogeneous structure of accretion and merger shocks,
the magnetic field distribution inside and outside CGs,
and the associated CR acceleration and propagation processes.
The CR confinement properties are especially crucial.
Although somewhat extreme,
if CR confinement in CGs is more efficient than we have assumed,
the contribution to the CR flux would decrease while the relative neutrino flux would increase.
Confinement may possibly be effective even on larger scales such as filaments or superclusters,
in which case the CR contribution could increase relative to the neutrino flux,
as our Galaxy resides in the local supercluster. 


\acknowledgments
K.M. and S.I. thank the referee, T. Kitayama, R. Blandford,
F. Takahara and H. Takami for useful comments.
K.M. is supported by a JSPS fellowship.
S.I. is supported in part by 
Grants-in-Aid for Scientific Research from the Ministry of E.C.S.S.T. (MEXT) of Japan,
Nos. 19047004 and 19540283.
S.N. is likewise partially supported by Nos. 19104006, 19740139 and 19047004.
Support also comes from the Grant-in-Aid for the Global COE Program
"The Next Generation of Physics, Spun from Universality and  
Emergence" from MEXT.

\clearpage





\end{document}